\documentclass[aps,prd,showpacs,groupedaddress,nofootinbib]{revtex4}

\usepackage{color}
\usepackage{graphicx}
\usepackage{amsmath}
\begin{document}

\title{Three-Dimensional Reconstruction of Erythrocyte in the Capillary}

\author{Y. F. Fan$^{1}$\footnote{Corresponding authors\\ Email: tfyf@gipe.edu.cn and yubofan@buaa.edu.cn}, Y. B. Fan$^{2*}$,
Z. Y. Li$^{3}$, W. T. Lin$^{1}$, Y. Wei$^{1}$, X. Zhong$^{4}$, T. Newman$^{3}$, C. S. Lv$^{1}$ and Y. Z. Fan$^{1,3}$}
\affiliation{$^{1}$Center for Scientific Research, Guangzhou
Institute of Physical Education, Guangzhou 510500, P.R. China\\
$^{2}$Bioengineering Department, Beijing University of
Aeronautics and Astronautics, Beijing 100191, P.R. China\\
$^{3}$International School, Jinan University, Guangzhou 510632, P.R. China\\
$^{4}$The First Affiliated Hospital of Jinan University, Guangzhou 510632, P.R. China}

\date{\today}

\begin{abstract}
The dynamic analysis of erythrocyte deformability is used as an important means for early diagnosis of blood diseases and blood rheology. Yet no effective method is available in terms of three-dimensional reconstruction of erythrocytes in a capillary. In this study, ultrathin serial sections of skeletal muscle tissue are obtained from the ultramicrotome, the tomographic images of an erythrocyte in a capillary are captured by the transmission electron microscope, and then a method to position and restore is devised to demonstrate the physiological relationship between two adjacent tomographic images of an erythrocyte. Both the modeling and the physical verification reveal that this method is effective, which means that it can be used to make three-dimensional reconstruction of an erythrocyte in a capillary. An example of reconstructed deformation of erythrocyte based on the serial ultrathin sections is shown at the end of this paper.
\end{abstract}


\maketitle

\section{Introduction}

The electron microscope has opened the gate to observe the surface and internal structure of micro objects on a nano-scale \cite{1}. Using electron beam scanning samples via scanning electron microscopy \cite{2}, we obtain the surface structure of micro objects, e.g. the surface structure of the muscle fiber \cite{3,4}. An ultrathin section is scanned by the transmission electron microscopy (TEM) and the organelle's internal structure, such as the sarcomere cross-section, is obtained \cite{5,6}. The electron tomography (ET) has brought the possibility to reconstruct the ultrastructure of a cell \cite{7,8}. The thick filament (myosin) model serves as an example \cite{9}. But even with ET, the methods to make three-dimensional (3-D) reconstructions of cells, organelles, or cell ultrastructure of a size of a few microns are still not effective enough, due to its limited electron beam penetration range into the biological samples \cite{10}.

A 3-D model of an erythrocyte in a capillary serves as a premise for a dynamic analysis. The erythrocyte dynamics has proven useful in the early diagnosis and prevention, clinical treatment, and pharmacological efficacy analysis of many types of blood diseases, ischemic heart disease, and ischemic cerebrovascular disease \cite{11,12}, bringing a great implication to 3-D reconstruction of erythrocyte in a capillary. Unfortunately, the reality that the size of red cells in mammals is only on a micron scale \cite{13} has presented great difficulty in reconstructing 3-D models of erythrocytes in a capillary by using the existing techniques and methods.

The ultramicrotome can provide serial ultrathin sections of an erythrocyte in a capillary. The erythrocyte tomographic image (ETI) is obtained by TEM, the ETIs stacks are set and then the 3-D erythrocytes are reconstructed. Methodologically, it sounds feasible to reconstruct a few microns of erythrocytes within an accuracy of a few nanometers ETI, but no satisfactory results of erythrocyte reconstruction based on the ultrathin sections are available \cite{14}. The cryo-ultramicrotome has made it possible to avoid treatment to samples - no fixation, no dehydration, no embedding and no staining are required \cite{15,16}. The invention of a contained console has automated the serial sections being placed in a grid holder \cite{8}. However, when these factors that impose constraints on the 3-D reconstruction are removed, another barrier appears, i.e. after an erythrocyte is cut, how can the relationships be restored between the shape and structure of the original adjacent ETIs by a parameter, not by an artificial match?

We assume that some physical quantities from ETI be applied to restore the physiological relationship between adjacent ETIs. Superpose the center of mass (COM) of adjacent ETIs, locate the preliminary positioning and use the principal axis of inertia (PAI) of ETI to set the preliminary orientation. Set the distance between the corresponding points to the boundary of adjacent ETIs as an action. Fine tune (translation and rotation) one of adjacent ETIs. When there is the minimum amount of action, the physiological relationship of adjacent ETIs is developed. If the modeling, simulation and physical verification support our hypothesis, then we can make the 3-D reconstruction of an erythrocyte in a capillary.

\section{Materials and methods}
This study has been carried out according to the existing rules and regulations of our institute's Ethnic Committee.

\subsection{Spiral CT scan}
The test equipment, provided by Image Processing Center of Zhujiang Hospital, was Brilliance 64-slice Scanner by Philips, Netherlands. Scan settings were: frame bone tissue; power: 120kv; pixel size: $0.50 mm$; layer distance: $0.50 mm$. The scanning was conducted along both feet transect, from top to bottom.
\subsection{Ultrathin sections}

Ultramicrotome: UC6 ultramicrotome (Leica Microsystems, Vienna, Austria).
Serial sections were produced by Zhongshan School of Medicine of Sun Yat-sen University. Tissue: adult male rat femoral triceps; Fixing agent: osmium tetroxide; Dehydration, embedding, and staining: methods shown in literature \cite{20}; Size of embedded section: trimmed as long strip of $0.20mm¡Á0.25mm¡Á1mm$; Grid: single hole copper grid, supporting a film thickness of approximately $20nm$ with carbon-coated polyvinyl formal membrane; Thickness of tissue: $50nm$.

\subsection{Transmission electron microscopy}
Philips Tecnai-10 (Philips, Eindhoven, the Netherlands) at $60KV$ and	Morada camera system (Soft Imaging Systems, M¨¹nster, Germany) were provided by the Center for Scientific Research of Jinan University.

\subsection{Software}
The 3-D model was constructed by Mimics (Version $10$) and MATLAB (Version $7.8$), which were provided by the Key Laboratory of Biomechanics and Mechanobiology of Ministry of Education.

\subsection{Positioning}
When the erythrocyte is cut into finite serial ultrathin sections by a single axial plane, the ETI stacks are set and can be restored to erythrocyte 3-D image. The x-ray irradiation of the organism via a single axial plane can bring the ETIs. The ETI stacks are set, and the surface point cloud computes results in the 3-D solid of the organism \cite{21,22}. The erythrocyte serial sections are cut by the ultramicrotome and their ETIs are captured by TEM. The serial sections are also ETIs along the single axial plane. The operations of placing the sections in the grid holder and the metal grid ($3 mm$ diameter) in a specimen chamber interrupt the positioning relationship of ETIs. Then, the difference between the TEM and the CT images lies in the fact that the physiological relationship between two adjacent ETIs is intermingled.

When the parallel planes can translate and rotate around an axis, there should be a location positioning and an orientation positioning. The two adjacent parallel planes bring two relative motions - translation and rotation. Two adjacent ETIs form a parallel relationship, so we position and orient the serial sections respectively: compute the COM and the PAI of every ETI. The superposition of the COM of two adjacent ETIs and the application of PAI as the coordinate of ETI facilitate the location and orientation positioning. Whether or not the positioning, after being located and oriented, should be regarded as the physiological relationship needs further verification.

\subsection{Restoration}
The surface of an erythrocyte is continuous, which means that no matter which axis the segmentation is done along, the boundary of the adjacent ETI is continuous. We describe this continuity as the distance between corresponding points of the adjacent ETI boundaries. This includes three steps: 1) build up a point-to-point relationship between the adjacent ETI boundaries; 2) compute the distance between the corresponding points; 3) take the sum of the distance between the corresponding points as the action, translate one of two adjacent ETIs and then rotate it. When the action is minimal, the physiological relationship between the ETIs is established.

To establish the relationship between points-to-points relations of adjacent ETI boundaries involves the following steps: 1) Take the origin where the PAIs interact at the boundary through erythrocyte COM as the starting point, and number the sequence boundary points counterclockwise; 2) normalize the number of boundaries that make up the erythrocyte adjacent ETIs to have the same number of adjacent ETI boundaries.
Compute the distance between the corresponding points of the adjacent ETI boundaries, using the following equation:

\begin{displaymath}
d_{j}=\sqrt{\left(x_{ij}-x_{i(j+1)}\right)^{2}+\left(y_{ij}-y_{i(j+1)}\right)^{2}},
\end{displaymath}
where \emph{j} stands for ETI serial number, \emph{i} for the serial number of the ETI boundary and $(x_{ij},y_{ij})$ for the relative COM position of a point on the boundary.

To identify the physiological relationship between two adjacent ETIs, the following objective function is applied:

\begin{displaymath}
\underset{1\leq k\leq n}{\min(\psi(k))}=\sum^{n}_{k=1}\sqrt{\left(x_{ij}-x_{(i+k)(j+1)}\right)^{2}+\left(y_{ij}-y_{(i+k)(j+1)}\right)^{2}},
\end{displaymath}

where \emph{n} refers to the number of ETI boundary points and then calculate the number by percentage, i.e. $n=100$.

\section{Results and Discussion}
\subsection{Reconstruction of cylinder}
\begin{figure}[!ht]
\begin{center}
\begin{tabular}{cccc}
 \includegraphics[width=12.8cm]{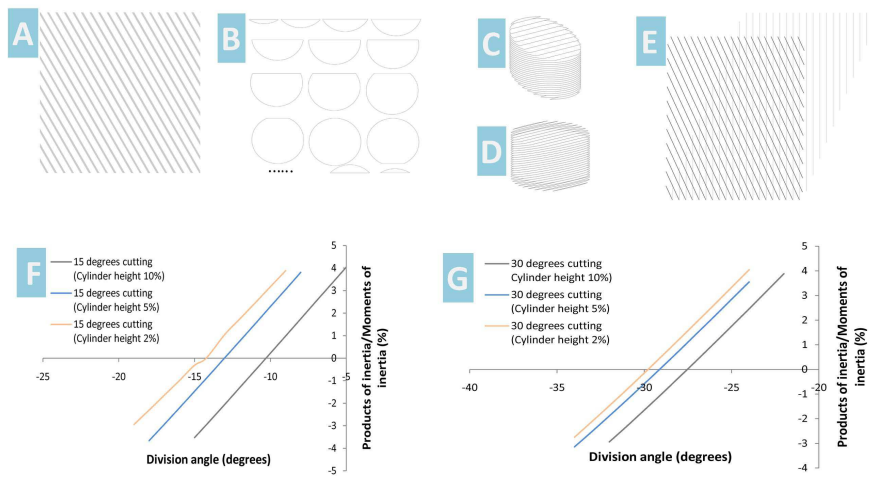}
\end{tabular}
\caption{\label{fig1} Reconstruction of a cylinder. Fig. 1A -B Divided cross sections along the long axis. Fig. 1C Positioning. Fig. 1D Restoration. Fig. 1E Cross sections before and after the restoration. Fig. 1F $15^{o}$ for divided cross section and the long axis. Fig. 1D $30^{o}$ for divided cross section and the long axis. In this Fig., the radius of the cylinder uses a unit of $1$ and so is the height. The cylinder is divided into $10$, $20$ and $50$ sections respectively along one unit of the long axis of the cylinder.}
\end{center}
\end{figure}

Bio-analytical method validation is necessary \cite{17}. To do so, we use modeling and simulation. First, use a homogeneous cylinder to examine the reconstruction of finite layers that are divided by a single axial plane (the long axis). See Fig. \ref{fig1}.

Fig. \ref{fig1} shows that after the homogeneous cylinder is divided into finite layers along the long axis, and when the layer distance is short enough, the boundary distance of the corresponding points can be taken as the least action to restore the original shape.

\subsection{Reconstruction of bone in vivo}
\begin{figure}[!ht]
\begin{center}
\begin{tabular}{cccc}
 \includegraphics[width=12.8cm]{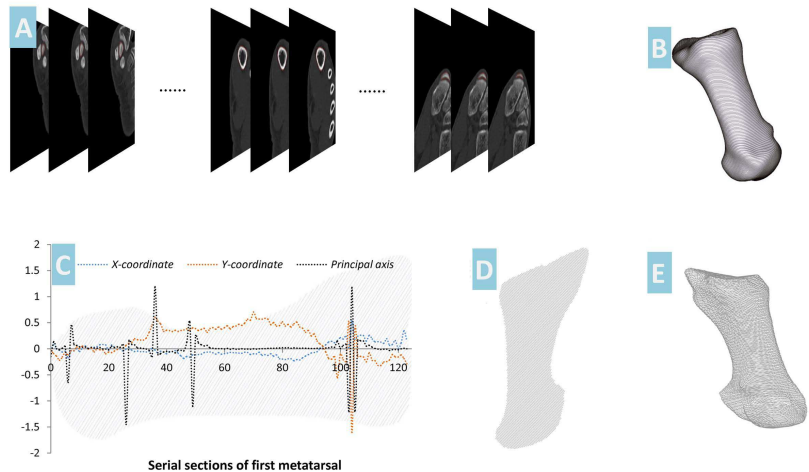}
\end{tabular}
\caption{\label{fig2} Reconstruction of the first phalanx ETIs. Fig. 2A Serial ETIs collected from CT. Fig. 2B Reconstruction result from reverse engineering software. Fig. 2C Relationship between COM and PAI of serial sections of first metatarsal. Fig. 2D-E Results from positioning and restoration. In Fig. 2C, the unit of axis $y$: position: $mm$; PAI: $radian$. }
\end{center}
\end{figure}

A cylinder is not completely symmetrical and its long axis is unique. Thus, using a cylinder to examine the effectiveness of the positioning and restoration has its limitations. A more reliable method is to use an asymmetrical object. In our experiment, we chose the human first phalanx. We collected the CT results of one subject \cite{18}. We translated and rotated the relationship between the adjacent first metatarsal images/ETIs randomly and observed the reconstruction results. See Fig. \ref{fig2}.

Fig. \ref{fig2} shows that when the relationship between the serial ETIs of the first phalanx is intermingled artificially, the 3-D images from positioning and restoration of the first phalanx are consistent with those from CT reconstruction. Our modeling and physical verification reveal that in a capillary, the deformed erythrocyte could be reconstructed.

\subsection{Reconstruction of Erythrocyte}	
\begin{figure}[!ht]
\begin{center}
\begin{tabular}{cccc}
 \includegraphics[width=12.8cm]{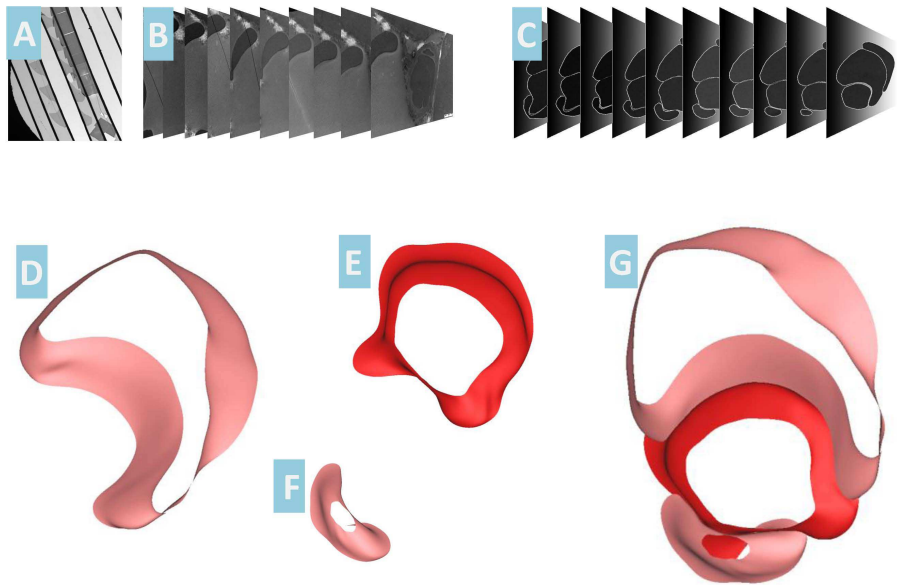}
\end{tabular}
\caption{\label{fig3} Relationship between serial ETIs. Fig. 3A Ultrathin serial sections of skeletal muscle tissue. Fig. 3B Serial sections of the capillary. Fig. 3C Serial sections of the erythrocyte. Fig. 3D-G Reconstructed erythrocyte in the capillary. In $400$ times, the muscle fiber shape is identified; in $2,000$ times, the capillary area; in $10,000$ times, the capillary; in $15,000$ times, the erythrocyte in a capillary to be used for reconstruction is identified.}
\end{center}
\end{figure}

We operate a sample along an axis of the ultramicrotome, cut it using a diamond knife and get serial sections of biological samples. The relationship between the adjacent sections can only be translation and rotation. The erythrocyte is extracted from the ETI and forms the erythrocyte serial sections. Superpose the COMs of adjacent sections, locate the preliminary positioning and use the PAI of ETM to set the preliminary orientation. Then we restore the physiological relationship between the adjacent ETIs by positioning and orientation. Results are shown in Fig. \ref{fig3}.

Fig. \ref{fig3} shows that operations such as placing serial sections in a grid holder can change the relationship between adjacent serial sections. The application of positioning and restoring methods can develop the physiological relationship between two adjacent ETIs. If we use cryo-ultramicrotome and contained console, then there is no need to undergo treatment such as fixation, dehydration, embedding or staining for the biological samples \cite{15,16}, and the sections can be automatically placed in the grid. Accordingly, a 3-D model of an erythrocyte in a capillary can be fully reconstructed.

Since the sample sections were placed manually, we were unable to get a sufficient number of serial sections required to build a complete 3-D model of an erythrocyte in a capillary. This needs to be further explored in our future study.

\section{Conclusion}
This study shows that the significance of the two physical quantities of COM and PAI on the ETI is threefold: 1) the shape change of ETI can be described by COM and PAI. When the section distance is short enough, the COMs and PAIs of adjacent ETIs are very close. This tells that the methods to use COM to locate and to use PAI to orient are reliable; 2) take the COM as the origin of the coordinate system and the PAI as the vertical axis, identify the coordinate of erythrocyte boundary position and offer a method to sequence the boundary points; 3) take the distance between the corresponding points of the two adjacent ETI boundaries as an action, the COM and PAI as variables; the objective function is used to determine the least action.

Our modeling and simulation results verify that our method to reconstruct the positioning and restoration of the erythrocyte in a capillary based upon the serial sections is feasible. The cryo-ultramicrotome can free us from treatments such as fixation, dehydration, embedding or staining for the biological samples \cite{15,16} while the contained console can place the sections in the grid automatically \cite{8}. Physical verification can help analyze the deformed erythrocyte in vivo in a capillary. In a micron-scale area, in addition to erythrocyte, there are important ultrastructures of sarcomere and mitochondria.

When a physical process takes place, nature always favors the minimal quantities \cite{19}. Erythrocyte deformability is a physical process. Set the distance between the corresponding points to the boundary of adjacent ETIs as an action. When the action minimizes, the physiological relationship between the adjacent ETIs is developed.

\section*{Acknowledgments}
This project was funded by National Natural Science Foundation of China under the Grant Numbers of $10925208$, $10972061$, $11172073$ and by Guangdong Natural Science Foundation under the grant numbers of $S2011010001829$. The authors would like to acknowledge the support from the subjects, Image Processing Center of Zhujiang Hospital, Zhongshan School of Medicine of Sun Yat-sen University, and Center for Scientific Research of Jinan University.

\end{document}